\documentclass[11pt]{article}
\usepackage[centertags]{amsmath}
\usepackage{amsfonts}
\usepackage{newlfont}
\usepackage{graphicx}
\usepackage{epstopdf}

\newtheorem{proposition}{Proposition}[section]
\newtheorem{remark}{Remark}[section]

\newtheorem{lemma}{Lemma}[section]

\def\sqr#1#2{{\vcenter{\vbox{\hrule height.#2pt
              \hbox{\vrule width.#2pt height#1pt \kern#1pt \vrule width.#2pt}
              \hrule height.#2pt}}}}
\def\signed #1{{\unskip\nobreak\hfil\penalty50
              \hskip2em\hbox{}\nobreak\hfil#1
              \parfillskip=0pt \finalhyphendemerits=0 \par}}
\def\endpf{\signed {$\sqr69$}}
\def\ms{\medskip}
\def\ds{\displaystyle}

\def\cF{{\cal F}}
\def\cG{{\cal G}}
\def\cH{{\cal H}}

\def\cN{{\cal N}}

\newcommand{\E}{\mathbf{E}}

\title{Random Time Forward Starting Options}
\author{F. Antonelli*,  A. Ramponi**, S. Scarlatti**{\footnote{e-mail: sergio.scarlatti@economia.uniroma2.it}} \\
DISIM, University of  L'Aquila - Italy \\
Dept. of Economics and Finance, University of Roma - Tor Vergata, Italy}

\date{March 2015}

\begin{document}

\maketitle
\begin{abstract}
We introduce a natural generalization of the  forward-starting options, first discussed by M. Rubinstein (\cite{Rubin}). The main feature of the contract presented here is that the strike-determination time is not fixed ex-ante, but  allowed to be random,  usually related to the occurrence of some event, either of financial nature or not. We will call these options {\bf Random Time Forward Starting (RTFS)}.

We show that, under  an appropriate  ``martingale preserving" hypothesis, we can exhibit arbitrage free prices, which can be explicitly computed in many classical market models, at least under independence between the random time and the assets' prices.  Practical implementations of the pricing methodologies are also provided.
Finally a credit value adjustment formula for these OTC options is computed for the unilateral counterparty credit risk.

\medskip

\noindent \textbf{Keywords}: Random times, forward-starting options, CVA.

\medskip
\noindent \textbf{JEL Classification}: G13

\end{abstract}

\section{Introduction}
\label{sec0a}

Forward-starting options are path dependent  put/call financial contracts
characterized by having a strike price expressed in terms of a
pre-specified percentage $\alpha$ of the asset price taken at a contractually
fixed  intermediate date $u\in (t,T]$, $T$ being the option maturity. The time
$u$ is known as {\it{strike-determination time}}.
The payoff of a forward starting call is therefore
\begin{equation}
\label{FSpay}
( S_T- \alpha S_u)^+.
\end{equation}
These products represent the fundamental component of
the  so-called cliquets (see \cite{GW99}), which  are indeed equivalent to a series of forward starting at-the-money options,
activated along a sequence of intermediate dates, upon payment of an initial  premium. Cliquets are often employed to buy protection against downside risk, though preserving an upside potential, for instance in pension plans in order  to hedge the guarantees attached to embedded equity linked products.

Wilmott in \cite{Wilmott} showed that these products are particularly sensitive to the model that one chooses for the dynamics of the underlying's price.

 In this paper we study a generalization of forward starting options allowing for random strike-determination times. That is to say, we hypothesize that we are making the forward starting option worth, if  some  future  random event  happens during the life of the contract with positive probability.
The value of the asset at this random time is then compared at maturity with the final value of the underlying, so the call's payoff  (\ref{FSpay}) is replaced by
\begin{equation}
\label{RTFSpay}
( S_T- \alpha S_{\tau\wedge T})^+,
\end{equation}
for some appropriate random time $\tau$, where we denoted by $x\wedge y =\min(x,y)$.
The pricing methodology  we will present is an extension of  Rubinstein's approach (see \cite{Rubin}),  and it can be applied to several existing option models.

It is easy to imagine a promising application of this type of contracts in several situations, as we show with the following examples.

\begin{itemize}
\item
 A  first example with a speculative goal,  could be represented by a call written (with $\alpha=1$) on the Manchester United stock (NYSE:{\bf{MANU}}), starting the beginning day of the Premiere League (PL) soccer tournament and ending the closing day of the tournament. The call is triggered by the first time, $\tau$, the United will eventually  lead the tournament with at least three points more than all other teams. If this event   happens, then the United will increase its chances to take part to the next Champions League (CL) tournament, an event which generates substantial additional cash-flows. If, at the end of PL tournament, the United will get the right to play CL, the share value $S_T$ of {\bf{MANU}} at that date  will probably  be
larger than $S_{\tau}$ and an investor in the call contract could pocket the difference.

\item  Another possible context we may imagine is when a firm expects
an  adverse  market scenario that might cause a downgrading of the firm's rating in the near future. So the firm's owner offers a contract of this type  to a smart manager (hired in the past), as an incentive  to convince him/her to remain within the firm and to work to improve the firm's position before maturity. As an example,  the firm's owner offers this contract to the manager on January, 1st 2015 with maturity $T=1$ year (with $\alpha=1$). If the event $\{\tau< T\}$ happens, for instance on June, 6th 2015, then the manager has time until maturity to make the firm's value rise again above $S_\tau$ and pocket the payoff, otherwise no reward is gained.
 The manager is therefore interested in assessing the correct current value of the incentive.

In this context, it is also interesting to consider the counterparty credit risk aspects of this type of contract from the point of view of the manager: if the firm (the counterparty) does not recover from the adverse market scenario, it might default a short time after the downgrading event or  default  could even    happen before downgrading. This leads to the problem of computing unilateral Credit Value Adjustment ({\bf CVA}) for this kind of option, since  the manager needs to evaluate this further risk  of the contract in order to assess correctly  the value of the firm's offer.

\item Lastly, it is also possible to consider contracts depending on more than one single random event: for example a cliquet call option could be triggered by the first time  the underlying asset reaches a pre-specified upper barrier $K_1$ until the asset reaches, if before maturity $T$, a new barrier level $K_2>K_1$, at that time  a second call option with strike $K_2$ and expiring at maturity gets activated. In general, depending on the characteristics of the financial product, one can imagine to construct  cliquet type options along  some increasing sequence of random times.

\end{itemize}

The literature on classical forward starting options is wide in terms of underlying's dynamics model and evaluation techniques. We shortly review the main contributions, since we will use these results as starting point in our analysis

Rubinstein (1991) (see \cite{Rubin})  was the first to consider the pricing problem of a forward-start
option in the  Black and Scholes model.

Later, valuation of forward starting options in stochastic volatility models was addressed by several authors.
Lucic (2003)  (see \cite{Lucic03})  wrote an evaluation formula by employing the asset's price at   determination time as numeraire, transforming those options in plain vanilla ones, giving a closed formula  for  the Heston's model. Kruse and N\"{o}gel (2005) (see \cite{Kruse05}), followed the same approach, but  they developed a pricing  formula exploiting the distributional properties of the Heston's model; they remarked the importance of such  an extension (w.r.t. the   Black-Scholes environment) since forward starting options are particularly sensitive to volatility changes.

Further,  Guo and Hung (2008) (see \cite{GH08}) derived  quasi-analytic pricing formulas   in general stochastic volatility/stochastic interest rates models, presenting jumps in both the asset's price and the volatility.  This was important, since, in contrast to the plain vanilla case, the value of a forward-starting option may not always increase with maturity. It depends on the current term structure of interest rates.
In the same context,  Ahlip and Rutkowski  (2009) (see \cite{AR}) chose a Heston's stochastic volatility model and a CIR dynamics for the  interest rates, with volatility and short rate both correlated with the dynamics of the stock returns. The main result is an analytic formula for the  call case,  derived by  using the probabilistic approach combined with the Fourier inversion technique, as developed in Carr and Madan (1999) (see \cite{CM}).
Van Haastrecht and  Pelsser (2011),  (\cite{HP}), exploiting  Fourier inversion of characteristic functions, developed a quantitative analysis of the pricing of forward starting options under stochastic volatility and stochastic interest rates (with Ornstein-Uhlenbeck dynamics), confirming that these  not only depend on future smiles, but are also very sensitive to model specifications such as volatility, interest rate and correlation movements, concluding  that it is of crucial importance to take all these factors into account for a correct valuation and risk management of these securities.
More recently, Ramponi (2012) (see \cite {Ram12}) extended this Fourier analysis  to regime switching models for the asset price.

Finally, in  2006 Chicago Board Options Exchange  introduced options on its implied volatility index (VIX), enhancing the interest in  forward starting options to forecast future volatility.
Albanese,  Lo and Mijatovic (\cite{AM09}) proposed a spectral  method  to evaluate numerically the joint distribution between the stock price and its realized variance, which gave  a way of pricing consistently European, general accrued variance payoffs, forward-starting and VIX options.

To evaluate our generalization of FS  options to random determination time, we  exploit
the techniques  developed in a number of papers (refer to  \cite{BJR} and to \cite{BR} for a general exposition)  to deal with default time in credit derivatives.
The main tool we employ  is the so called  assumption (H),  that  guarantees that all martingales remain as such when adding the information generated by the default time. In other words, this means that the no-arbitrage structure is not altered by the introduction of the random time. This leads to the so called key lemma, that allows to switch from the global information  (generated by the asset's price and the random time) to the information generated only by the observable prices.

Following the intensity based approach, we give a general expression of the pricing formula  either in conditions of independence or not. Under independence between the asset's price and the random time,    we specialize and explicit our prices in several computable models  (Black \& Scholes,  Heston, Variance Gamma) and we suggest an affine dependency structure for the intensity when the independence assumption  is dropped,  showing  it is still possible to recover a partially easily computable formula.

Finally, keeping in mind  the second example presented above, we focus on computing the Credit Value Adjustment,  due to counterparty risk, required by these products, when in presence of a further default time (for a general reference  see  Brigo and Chourdakis (2009), \cite {BC}).

The paper is organized as follows.  Section \ref{sec1a} gives the set up, in Section \ref{sec2a} we  present a general approach to the evaluation of this kind of product, in  Section \ref{sec3a} we develop some explicitly computable cases under independence. Section \ref{sec4a} provides some numerical implementation of the derived pricing formulas, finally  Section \ref{sec5a} treats briefly the counterparty risk implied by this product, when a default time for the issuer might  occur.

\section{The set up}
\label{sec1a}

We consider a time interval $[0,T]$, with $T$  denoting the maturity of the contracts.
We assume that the market is made up of a single asset and a  bond  living in a probability space $(\tilde \Omega, \cal F, P)$,   with a filtration $\{\cal F_s\}_{s\geq 0}$,   made complete and right continuous to satisfy the {\it usual hypotheses} (see \cite{P}). Moreover, we assume we are in absence of arbitrage opportunity and that $P$ is a risk neutral measure selected by some criterion.

Before formalizing mathematically the new contract previously described, we need to recall briefly
the evaluation of classical forward-start options.

Denoting by  $B(t,T)$ the discounting process,  the price of a forward-start call with percentage $\alpha$ and  strike-determination time $u>t$ is given by
\begin{equation}\label{deterministic}
c(t,u,T)=\E^P[B(t,T)(S_T-\alpha S_{u})^+|\cF_t],\qquad 0\le t\le  T.
\end{equation}
In market models where the price of a plain vanilla call option is represented by a deterministic homogeneous function of degree $1$ in two spatial variables $(x,y)$, that is
$$
call(t,\gamma x, \gamma y, T) = \gamma call(t,x,y, T), \qquad   \forall \,\gamma, x,y >0
$$
applied to the given pair $(S_t,K)$, the price of a classical (call) forward-start option  is
\begin{eqnarray}
\nonumber
c(t,u,T)\!\!\!&=&\!\!\!\E^P[B(t,u)\E^P[B(u,T)(S_T-\alpha S_{u})^+|\cF_u]|\cF_t]=\E^P[B(t,u) call(u,S_u,S_u\alpha, T)|\cF_t]\\
\nonumber\!\!\!&=&\!\!\!\E^P[B(t,u)call(u, \gamma 1, \gamma \alpha , T)\Big |_{\gamma=S_u} |\cF_t]=\E^P[B(t,u)(\gamma call(u, 1, \alpha , T))\Big |_{\gamma=S_u} |\cF_t]\\
\label{hom}\!\!\!&=&\!\!\!
\E^P[B(t,u)S_u call(u,1,\alpha,T) |\cF_t].
\end{eqnarray}
Scale-invariant models (see \cite{AN07}) certainly hold this property. Furthermore, if the call price is a deterministic function of the current price of the underlying and it does not include any additional stochastic factor,   we may conclude  that  the price of the forward-start option verifies
\begin{equation}
\label{scale}
\E^P[B(t,u)S_u call(u,1,\alpha,T)|\cF_t] =
\E^P[B(t,u)S_u|\cF_t] call(u,1,\alpha,T) = S_t \
call(u,1,\alpha, T),
\end{equation}
where in the last equality we used the martingale property of the discounted asset price. Notice that in such a case the portfolio having $call(u,1,\alpha, T)$ units of the stock `replicates" the $t$-value of the forward-start
option.

Sometimes an analogous result may be achieved also in presence of stochastic factors via change of numeraire arriving at a formula of the type
\begin{equation}
\label{chscale}
\E^P[B(t,u)S_u call(u,1,\alpha,T)|\cF_t] =S_t\E^{\bar P}[call(u,1,\alpha,T)|\cF_t]
\end{equation}
for some proper equivalent measure $\bar P$.

We  want  to generalize these contracts allowing  the strike-determination time to be random, thus we need to consider a market model (and hence a probability space) where  the asset price $S$ lives together with a random time $\tau$. We therefore  assume that there exists a $\sigma-$algebra $\cG$ containing $\cF$ and a filtration $\{\cal G_s\}_{s\ge 0}\subseteq \cG$, satisfying the usual hypotheses,  that makes $S_t$ and $H_t= \mathbf 1_{\{\tau\le t\}}$ adapted processes (that is to say that $\tau$ is stopping time w.r.t.  $\{\cal G_s\}_{s\ge 0}$ ) and that we may extend the probability $P$ to a new probability measure $Q$ on the  measure space  $( \Omega, \cal G)$. Hence, referring to the previous notation, we have
$
\cal F_t \subseteq \cal G_t, $ for all  $ t \ge 0$ and  from now on we will assume
$
\cal G_t=\cal F_t \lor \cal H_t,
$
where $\cal H_t= \sigma (H_s\,:\,s\le t)$.
We recall that the filtration stopped at the stopping time $\tau$ will be denoted $\cG_\tau$ and it means
$$
\cG_\tau=\{ A \in \cG\, :\,  A\cap \{\tau\le t\}\in \cG_t, \, \forall \, t \ge 0\}.
$$
Besides we introduce the classical and fundamental hypothesis

\medskip

\noindent
{\bf (H)}\qquad \qquad\qquad \qquad  Every $\cF_t-$martingale remains a $\cG_t-$martingale.\hfill

\medskip
This assumption, known as the  H-hypothesis, is widely used  in the credit risk literature (see e.g. \cite{GJ08} and references therein). With this hypothesis, we may conclude that $S_u$ remains in general a $\cG_u-$semimartingale and $B(t,u)S_u$ a martingale under the extended probability $Q$. So we can replace the arbitrage free pricing formula (\ref{deterministic}) by the more general expression
\begin{equation}\label{random}
c(t,T)=\E^{Q}[B(t,T)(S_T-\alpha S_{\tau\wedge T})^+|\cal G_t].
\end{equation}

From now on, this product will be called  a {\bf Random Time Forward Starting } option and we will refer to it as {\bf RTFS} options. Notice that the payoff remains $\cG_T-$measurable. In what follows we treat only the call  case as, under hypothesis (H),  a natural extension of the put-call parity holds   for these products
$$
\E^{Q}[B(t,T)(S_T-\alpha S_{\tau\wedge T})^+|\cal G_t]- \E^{Q}[B(t,T)(\alpha S_{\tau\wedge T}- S_T)^+|\cal G_t]
= S_t - \alpha S_{\tau\wedge t} B(t, \tau\wedge t).
$$

Lastly we remark that for  $\alpha=1$, these options are worth something
only if the random occurrence happens before maturity, in a sense we could say they are  triggered by the random time.

\section{Pricing formulas}
\label{sec2a}

In this section we want to understand how to make formula (\ref{random}) more explicitly computable.

As a general consideration, let us notice that
\begin{eqnarray}
\label{general}
\nonumber c(t,T)&=&
 \E^{Q}[B(t,T)(S_T-\alpha S_{\tau\wedge T})^+(\mathbf 1_{\{0<\tau\le T\}}+ \mathbf 1_{\{ \tau>T\}})|\cal G_t]\\
\nonumber &=& \E^{Q}[B(t,T)(S_T-\alpha S_{\tau})^+\mathbf 1_{\{0< \tau\le T\}}|\cal G_t]+(1-\alpha) \E^{Q}[B(t,T)S_T \mathbf 1_{\{ \tau>T\}}|\cal G_t]\\
\nonumber &=& \E^{Q}[B(t,T)(S_T-\alpha S_{\tau})^+|\cal G_t]\mathbf 1_{\{0< \tau\le t\}}+
\E^{Q}[B(t,T)(S_T-\alpha S_{\tau})^+\mathbf 1_{\{t< \tau\le T\}}|\cal G_t]\\
&+&(1-\alpha)S_t\mathbf 1_{\{t<\tau\}}- (1-\alpha)\E^{Q}[B(t,T)S_T\mathbf 1_{\{ t< \tau\le T\}})|\cal G_t].
\end{eqnarray}
  having written $\mathbf 1_{\{\tau>T\}} $ as $1- \mathbf 1_{\{ 0< \tau\le T\}}$.

For the first term there is not much to say, there $\tau\le t $   and it behaves as a call price with strike price $\alpha S_\tau$, which is completely known at time $t$. The third term represents the guaranteed payoff of this contract, so we have to study the remaining terms. Hypothesis (H)
is going to help us for the last one. We know that  $B(t, u)S_u$ is a martingale under $\cG$ and that  the event $\{t< \tau\le T\}\in \cG_\tau$ (since  for any $ t\le u\le T $, $\{t< \tau\le T\}\cap \{ \tau\le u\}= \{t< \tau\le u\}\in \cG_u$), so conditioning internally w.r.t. $\cG_\tau $, we obtain
\begin{eqnarray*}
&&\E^{Q}[B(t,T)S_T\mathbf 1_{\{ t< \tau\le T\}})|\cal G_t]=\E^{Q}[\E^Q(B(t,T)S_T\mathbf 1_{\{ t< \tau\le T\}}|\cG_\tau)|\cal G_t]\\
&=&\E^{Q}[\E^{Q}(B(t,T)S_T|\cal G_\tau)\mathbf 1_{\{ t< \tau\le T\}}|\cal G_t]=\E^{Q}(B(t,\tau)S_\tau\mathbf 1_{\{ t< \tau\le T\}}|\cal G_t],
\end{eqnarray*}
where we used the optional sampling theorem, assuming enough integrability of the asset price process, as $\tau\le T$ almost surely.

The second term  may be rewritten as
\begin{eqnarray*}
&&\E^{Q}[B(t,T)(S_T-\alpha S_{\tau})^+\mathbf 1_{\{ t< \tau\le T\}})|\cal G_t]=\E^{Q}[\E^Q(B(t,T)(S_T-\alpha S_{\tau})^+\mathbf 1_{\{ t< \tau\le T\}}|\cG_\tau)|\cal G_t]\\
&=&\E^{Q}[B(t,\tau)\E^{Q}(B(\tau, T)(S_T-\alpha S_{\tau})^+|\cal G_\tau)\mathbf 1_{\{ t< \tau\le T\}}|\cal G_t]=\E^Q[B(t,\tau) c(\tau,\tau, T)\mathbf 1_{\{ t< \tau\le T\}}|\cal G_t],
\end{eqnarray*}
with  $c$  defined w.r.t. $\cG_t$ as in (\ref{deterministic}), where we extended the definition to include also $u=t$.

\medskip
To proceed with our evaluation we have two possibilities. Either $\cG_t\equiv\cF_t$, i.e. $\tau $ is   an $\cF_t-$stopping time, or not. In the first case, some computable situations may occur. For instance, if
 $\tau=\inf\{u>0 : S_u\ge H\}$ is   a hitting time (choosing the barrier level $H>S_0$), we have
\begin{eqnarray*}
c(t,T)
&=& \E^{Q}[B(t,T)(S_T-\alpha H)^+|\cF_t]\mathbf 1_{\{0< \tau \le t\}}+
\E^{Q}[B(t,T)(S_T-\alpha H)^+\mathbf 1_{\{t< \tau \le T\}}|\cF_t]\\
&+&(1-\alpha)S_t\mathbf 1_{\{t<\tau\}}- (1-\alpha)\E^{Q}[B(t,T)S_T\mathbf 1_{\{ t< \tau\le T\}})|\cF_t].
\end{eqnarray*}
Summarizing
\begin{eqnarray}
c(t,T)&=&\textrm{call price}(t,T, S_t, \alpha H)\mathbf 1_{\{0< \tau\le t\}}+\textrm{barrier price}^{up/in}(t,T, S_t, \alpha H, H)\\
\nonumber
&-& (1-\alpha)\E^{Q}[B(t,T)S_T\mathbf 1_{\{ t< \tau\le T\}})|\cF_t]+(1-\alpha)S_t\mathbf 1_{\{t<\tau\}},
\end{eqnarray}
where ``barrier price" denotes the price of a barrier call option
with strike price $\alpha H$ which is activated as soon as the
threshold $H$ is reached by the asset price.
If this event does not happen before maturity $T$, at maturity the
contract is not going to pay zero to the holder as in standard barrier contracts, but
$(1-\alpha)$-percent of the terminal asset price $S_T$, since the first three pieces of the  formula are zero.
Therefore, if $\tau>T$, this random starting  forward option differs from
 a plain vanilla  barrier option by a specific  non negative value,  never exceeding $(1-\alpha)H$.

Other situations may be more or less complex to evaluate, depending upon the definition of the stopping time $\tau$.

\medskip

In general, since the observable is the asset's price process,   it would be interesting to be able to write the pricing formula in terms of $\cF_t$, rather than
 ${\cal G}_t={\cal F}_t \vee {\cal H}_t$.
For that we have  the following key Lemma (see \cite{BJR}, \cite{BCB}).

\begin{lemma}\label{key} For any integrable $\cG-$measurable r.v. $Y$, the following equality holds
\begin{equation}
\label{key}
\E^Q\Big[\mathbf 1_{\{\tau>
t\}}Y|\cG_t\Big]=Q(\tau>t|\cG_t)\frac{\E^Q\Big[\mathbf 1_{\{\tau>
t\}}Y|\cF_t\Big]}{Q(\tau>t|\cF_t)}.
\end{equation}
\end{lemma}

Applying this lemma to the second and fourth term of (\ref{general}), respectively with
$$
Y=B(t,\tau)c(\tau, \tau, T)\mathbf 1_{\{t< \tau\le T\}}\qquad \textrm{
and}\qquad
Y=B(t,\tau)S_\tau\mathbf 1_{\{t< \tau\le T\}}
$$
and remembering that $1-H_t=\mathbf 1_{\{\tau>t\}}$ is $\cG_t-$measurable, we obtain
\begin{eqnarray}
\label{key1}
\E^Q[B(t,\tau)c(\tau,\tau, T)\mathbf 1_{\{t< \tau\le T\}}|\cal G_t]&=&\mathbf 1_{\{\tau>t\}}
\frac{\E^Q[B(t,\tau)c(\tau,\tau, T)\mathbf 1_{\{t< \tau\le T\}}|\cF_t]}{Q(\tau>t|\cF_t)}\\
\label{key2}
\E^Q[B(t,\tau)S_\tau\mathbf 1_{\{t< \tau\le T\}}|\cal G_t]&=&\mathbf 1_{\{\tau>t\}}
\frac{\E^Q[B(t,\tau)S_\tau\mathbf 1_{\{t< \tau\le T\}}|\cF_t]}{Q(\tau>t|\cF_t)}.
\end{eqnarray}
The above may be rewritten following the hazard process  approach. Let us denote the conditional distribution of the default time  $\tau$ given $\cF_t$ as
$$
F_t=Q(\tau\leq t|\cF_t), \qquad \forall \, t\ge 0
$$
 and let us notice that for $u\ge t$,  $Q(\tau\leq u|\cF_t)=\E^Q(Q(\tau\leq u|\cF_u)|\cF_t)=\E^Q(F_u|\cF_t)$.

In order to apply the so called intensity based approach,   we need to assume that
$F_t(\omega)<1$ for all  $t>0$ (which automatically excludes  that $\cG_t\equiv \cF_t$) to well define the  so called risk or hazard process
\begin{equation}
\label{risk}
 \Gamma_t:=-\ln(1-F_t)\quad\Rightarrow \quad F_t= 1 - \mathrm e^{-\Gamma_t}\quad \forall \, t>0, \qquad  \Gamma_0=0.
\end{equation}
With this notation, we rewrite (\ref{key1}) and (\ref{key2}) as
\begin{eqnarray}
\label{key3}
&&\E^Q[B(t,\tau)c(\tau,\tau, T)\mathbf 1_{\{t< \tau\le T\}}|\cal G_t]=(1-H_t)\E^Q[B(t,\tau)c(\tau,\tau, T)(H_T-H_t)\mathrm e^{\Gamma_t}|\cF_t]\\
\label{key4}
&&\E^Q[B(t,\tau)S_\tau\mathbf 1_{\{t< \tau\le T\}}|\cal G_t=(1-H_t)\E^Q[B(t,\tau)S_\tau(H_T-H_t)\mathrm e^{\Gamma_t}|\cF_t],
\end{eqnarray}
but by proposition 5.1.1 (ii) page 147 of \cite{BR}  these conditional expectations may be written as
\begin{eqnarray}
\label{key5}
\E^Q[B(t,\tau)c(\tau,\tau, T)(H_T-H_t)|\cF_t]&=&\E^Q[\int_t^T B(t,u)c(u,u, T) dF_u|\cF_t]\\
\label{key6}
\E^Q[B(t,\tau)S_\tau(H_T-H_t)|\cF_t]&=&\E^Q[\int_t^T B(t,u)S_udF_u|\cF_t].
\end{eqnarray}
If we have the additional hypotheses  ${\cal G}_t\equiv {\cal F}_t \vee {\cal H}_t= \cal F_t \otimes  \cal H_t$ for all $t$ and
\begin{description}
\item[(IND)] \hspace{3.5cm} $\cal F$ and $\cal H$ are independent,
\end{description}
then $\tau $ is independent of $\cF_t$, hypothesis (H) is automatically satisfied and $F_t=Q(\tau\le t)$ is deterministic,
so the previous formula becomes
\begin{eqnarray}
\label{gener}
\nonumber
&&\E^Q[B(t,\tau)c(\tau,\tau, T)(H_T-H_t)|\cF_t]=\E^Q[\int_t^T B(t,u)c(u,u, T) dF_u|\cF_t]\\
\nonumber&=&\int_t^T\E^Q[ \E^Q(B(t,T)(S_T-\alpha S_u)^+|\cF_u)|\cF_t]dF_u=\int_t^T\E^Q[ B(t,T)(S_T-\alpha S_u)^+|\cF_t] dF_u\\
&=&\int_t^Tc(t,u,T)dF_u
\end{eqnarray}
and the other is treated similarly.

We can summarize the above results in the following

\begin{proposition}\label{gener1} With the above notation, we have
\begin{enumerate}
\item[(a)] Under hypothesis $(H)$ the price of a {\bf RTFS} call option is given by
\begin{eqnarray*}
c(t,T)&=& \E^{Q}[B(t,T)(S_T-\alpha S_{\tau})^+|\cal G_t]\mathbf 1_{\{0< \tau\le t\}}+(1-\alpha)S_t\mathbf 1_{\{\tau>t\}}\\
&+&\mathbf 1_{\{\tau>t\}}\E^Q[\int_t^T B(t,u)c(u,u, T)\mathrm e^{-(\Gamma_u-\Gamma_t)}d\Gamma_u|\cF_t]\\
&-&(1-\alpha)\mathbf 1_{\{\tau>t\}}
\E^Q[\int_t^T B(t,u)S_u \mathrm e^{-(\Gamma_u- \Gamma_t)}d\Gamma_u|\cF_t]
\end{eqnarray*}
\item[(b)] Under hypothesis $(IND)$ the price of a {\bf RTFS} call option is given by
\begin{eqnarray*}
c(t,T)
&=& \E^{Q}[B(t,T)(S_T-\alpha S_{\tau})^+|\cal G_t]\mathbf 1_{\{0< \tau\le t\}}\\
&+&\mathbf 1_{\{\tau>t\}}\Big [\int_t^Tc(t,u, T)\mathrm e^{-(\Gamma_u- \Gamma_t)}d\Gamma_u+(1-\alpha)S_t\mathrm e^{-(\Gamma_T-\Gamma_t)}
\Big ]
\end{eqnarray*}
\end{enumerate}

\end{proposition}

\noindent
\begin{Proof} Part (a) is only the assembling of the various pieces previously presented. As for part (b), because of independence, $\Gamma $ is deterministic and it gets pulled out of the expectation. Since $Q$ is a martingale measure,
the integrands of the last two terms verify the martingale property w.r.t. $\cF_t$ under $Q$.
\endpf
\end{Proof}

\begin{remark}\label{rem1}
If we assume that  $\Gamma_t$ is absolutely continuous  w.r.t. the Lebesgue measure, that is to say
\begin{equation}
\label{acrisk}
 \Gamma_t=\int_0^t \lambda_sds \qquad \forall  t>0
\end{equation}
for some  $\cF_t-$adapted and non negative process  $\lambda_t$ called intensity (see \cite{BJR}), then also
\begin{eqnarray}
 F_t=1-e^{-\Gamma_t}=1-e^{-\int_{0}^{t} \lambda_{s}ds}
\end{eqnarray} is absolutely continuous and if we denote by $f_t$  its density, necessarily we have
 $\displaystyle\lambda_t=\frac{f_t}{1-F_t}$

In this case  we  have
\begin{enumerate}
\item[(a)] Under hypothesis $(H)$ the price of a {\bf RTFS} call option is given by
\begin{eqnarray}
\label{hazard1}
c(t,T)&=& \E^{Q}[B(t,T)(S_T-\alpha S_{\tau})^+|\cal G_t]\mathbf 1_{\{0< \tau\le t\}}+(1-\alpha)S_t\mathbf 1_{\{\tau>t\}}\\
\nonumber &+&\mathbf 1_{\{\tau>t\}}\E^Q[\int_t^T B(t,u)c(u,u, T) \lambda_u\mathrm e^{-\int_t^u\lambda_s ds}du|\cF_t]\\
 \nonumber &-&(1-\alpha)\mathbf 1_{\{\tau>t\}}
\E^Q[\int_t^T B(t,u)S_u \lambda_u\mathrm e^{-\int_t^u\lambda_s ds}du|\cF_t]
\end{eqnarray}
\item[(b)] Under hypothesis $(IND)$ the price of a {\bf RTFS} call option is given by
\begin{eqnarray}
\label{FSprice_indip}
c(t,T)&=& \E^{Q}[B(t,T)(S_T-\alpha S_{\tau})^+|\cal G_t]\mathbf 1_{\{0< \tau\le t\}}\\
\nonumber&+&\mathbf 1_{\{\tau>t\}}\Big [\int_t^Tc(t,u, T) \lambda_u\mathrm e^{-\int_t^u\lambda_s ds}du+(1-\alpha)S_t\mathrm e^{-\int_t^T\lambda_s ds}
\Big ]
\end{eqnarray}
\end{enumerate}
The above formulas, either in presence of independence or not, rely on the knowledge of the model and of the conditional distribution of the  default time. Hence, deriving an explicit or implementable formula for the pricing will depend heavily on the modelling choices we are going to make.

Of course, the independent case is much easier than the other and we focus on it in the next section, to show that there are some interesting  treatable cases. We will always  take $\alpha=1$.
\end{remark}

{\remark
{\bf (Affine hazard rates)}

When  we  no longer assume independence of $\tau $ and $S$, we need to resort to  formula  (\ref {hazard1}) and a reasonable choice is to make the  hazard rate depend also on the price process. As a first attempt one may try an affine model, so we decide that
\begin{equation}
\label{affinelam}
\lambda_u= a(u) S_u+ b(u) Z_u,
\end{equation}
where $a, b$ are deterministic, positive and bounded functions, while $Z$ is a  positive $\cG_t-$adapted process independent of $\cF_t$, $0\le t\le T$.

By using the optional projection theorem (see \cite{DM} theorem VI.57),  we may arrive at
 the pricing formula
\begin{eqnarray*}
c(t,T)\!\!\!
&=&\!\!\! \E^{Q}[B(t,T)(S_T-\alpha S_{\tau})^+|\cal G_t]\mathbf 1_{\{0< \tau\le t\}}\\
\!\!\!&+&\!\!\!\int_t^T a(u)call(u,1,\alpha, T)\E^Q[  B(t,u)S_u^2\mathrm e^{-\int_t^u  a(s) S_sds}|\cF_t]\E^Q[\mathrm e^{-\int_t^u  b(s) Z_sds}]du\\
\!\!\!&+&\!\!\!\int_t^T b(u)call(u,1,\alpha, T)\E^Q[   B(t,u)S_u\mathrm e^{-\int_t^u  (a(s) S_sds}|\cF_t]\E^Q[Z_u\mathrm e^{-\int_t^u  b(s) Z_sds}]du\\
\!\!\!&+&\!\!\!(1-\alpha)\E^Q[B(t,T)S_T\mathrm e^{-\int_t^T a(s) S_sds}|\cF_t]\E[ \mathrm e^{-\int_t^T b(s) Z_sds}]\Big \},
\end{eqnarray*}
where all the above  expectations are of the same type. If not explicitly computable, the expectations can be evaluated via Monte Carlo  or transform methods, once a grid of points $t=u_0<u_1<\dots<u_N=T$ in order to approximate the time integrals has been fixed. More precisely, one applies  the Monte Carlo approach to the inner expectations  by generating $M$ independent paths of the processes $S$ and $Z$ at the grid points.}

\section{Some computable models}
\label{sec3a}

\ms
We now show some examples where the above formulas for the RTFS options can be explicitly
computed, of course the first case to analyze is the Black and Scholes model.

\bigskip
\noindent \textbf{Pricing in the Black \& Scholes market}
\ms

In the classical BS market with constant risk-free rate $r$ and volatility $\sigma$, the  price of a plain vanilla European call with maturity $T$ and strike $K$ will be denoted by $call^{BS}(t, S_t, K, T)$, while the price of of a forward-start option with maturity $T$, strike-determination time $u$ and percentage $\alpha=1$ (see \cite{Rubin}) is simply
\begin{equation} \label{FS_priceBS}
c^{BS}(t,u,T) =S_t call^{BS}(u, 1, 1,\sigma, r,T), \qquad u\ge t
\end{equation}
where
$$
call^{BS}(t,S_t,K; \sigma,r,T) = S_t \cN(d_1)- e^{-r (T-t)} K
\cN(d_2)
$$
\begin{eqnarray*}
d_1 = \frac{\log(S_t/K)+(r+\sigma^2/2)(T-t)}{\sigma \sqrt{T-t}}\qquad
d_2 = \frac{\log(S_t/K)+(r-\sigma^2/2)(T-t)}{\sigma \sqrt{T-t}},
\end{eqnarray*}
hence in formula (\ref {FS_priceBS}) we have
$
d_i=c_i
\sqrt{T-u}, \quad i=1,2 , \quad
c_1 =(\frac r\sigma  + \frac \sigma 2),\quad  c_2 = c_1 -\sigma
$.
Let us notice that $c_1^2>r$ is always true.

Henceforth, under (IND), the price of a ({\bf{RTFS}}) option is
given by
\begin{eqnarray}\label{GFSpriceBS}
c(t,T)= \mathbf 1_{\{0< \tau\le t\}}call^{BS}(t, S_t, S_\tau,\sigma, r,T)+\mathbf 1_{\{\tau>t\}}S_t\!\!\int_t^T \!\! call^{BS}(u, 1, 1,\sigma, r,T) \lambda_u\mathrm e^{-\int_t^u\lambda_s ds}du.
\end{eqnarray}
The valuation of  (\ref{GFSpriceBS}) can be found in closed form in some simple cases. If $\tau \sim
\exp(\lambda)$, $\lambda >0$ under $Q$, then  for the second term we obtain
\begin{equation} \label{BSform}
\mathbf 1_{\{\tau>t\}}S_t\int_t^Tcall^{BS}(u, 1, 1,\sigma, r,T) \lambda_u\mathrm e^{-\int_t^u\lambda_s ds}du=\mathbf 1_{\{\tau>t\}}S_t\Big [A_1(t,T)-A_2(t,T)\Big ],
\end{equation}
where
\begin{equation} \label{A1A2BS}
A_1(t,T)=\lambda\mathrm e^{\lambda t}\!\!\int_t^T\!\! \cN(c_1\sqrt{T-u})\mathrm e^{-\lambda
u}du,\,\,\,
A_2(t,T)=\lambda\mathrm e^{\lambda t}\!\!\int_t^T\!\!\mathrm e^{-r(T-u)}\cN(c_2\sqrt{T-u})\mathrm e^{-\lambda u}du
\end{equation}
and it is clear that $A_1>A_2$ for all values of the parameters.

The above integrals specialize depending on the choice of the parameters. If $r\neq \lambda$ we have three cases.

If $c_1^2>2\lambda$  then these integrals can be explicitly  computed exploiting integration by parts and the Gaussian density, obtaining (keeping in mind that  $c_2^2- 2(\lambda-r)=c_1^2-2\lambda$)
\begin{eqnarray}
\label{case1a1}
\!\!\!\!\!\!A_1(t,T)\!\!\! &=&\!\!\! \cN(c_1 \sqrt{T\!-t})- \frac{c_1\mathrm e^{-\lambda(T\!-t)} }{\sqrt{c_1^2\!-\!2\lambda}}\cN(\sqrt{(c_1^2\!-\!2\lambda)(T\!-t)})
+\frac {\mathrm e^{-\lambda(T\!-t)} }2(\frac{c_1}{\sqrt{c_1^2\!-\!2\lambda}}-1)\\
\label{case1a2}
\!\!\!\!\!\!A_2(t,T)\!\!\! &=&\!\!\! \frac \lambda{\lambda\!-\! r} \Big [\mathrm e^{-r(T\!-t)}\cN( c_2\sqrt{T\!-t})-   \frac{c_2\mathrm e^{-\lambda(T\!-t)}}{\sqrt{c_1^2\!-\!2\lambda}}\cN(\sqrt{(c_1^2\!-\!2\lambda)(T\!-t)})\\
\nonumber
&+&\frac {\mathrm e^{-\lambda(T-t)}}2(\frac{c_2}{\sqrt{c_1^2\!-\!2\lambda}}-1)\Big].
\end{eqnarray}
When $c_1^2=2\lambda$ (hence $c_1^2=2(\lambda-r)$), then the above formulas reduce to
\begin{eqnarray}
\label{case2a1}
A_1(t,T)\!\!\! &=&\!\!\! \cN(c_1 \sqrt{T-t})-\mathrm e^{-\lambda(T-t)}(\frac 12+  \frac{c_1\sqrt{T-t}}{\sqrt{2\pi}})\\
\label{case2a2}
A_2(t,T)\!\!\! &=&\!\!\! \frac \lambda{\lambda- r}\Big [\mathrm e^{-r(T-t)}\cN(c_2 \sqrt{T-t})-\mathrm e^{-\lambda(T-t)}(\frac 12+  \frac{c_2\sqrt{T-t}}{\sqrt{2\pi}})\Big ]
\end{eqnarray}
and for $c_1^2<2\lambda$ we lose the Gaussian integral and  we arrive at
\begin{eqnarray}
\label{case3a1}
\!\!\!\!\!\!A_1(t,T)\!\!\! &=&\!\!\! \cN(c_1 \sqrt{T-t})-\mathrm e^{-\lambda(T-t)} \Big [ \frac 12 +\frac{c_1}{\sqrt{2\lambda-c_1^2}}
\int^{\sqrt{(2\lambda- c_1^2)(T-t)}}_0 \frac{\mathrm e^{\frac{z^2}2}}{\sqrt{2\pi}}dz\Big ]\\
\label{case3a2}
\!\!\!\!\!\!A_2(t,T)\!\!\! &=&\!\!\! \frac \lambda{\lambda\!-\! r} \Big \{\mathrm e^{-r(T\!-t)}\cN( c_2\sqrt{T\!\!-\!t})\!- \!\mathrm e^{-\lambda(T\!-t)} \Big [ \frac 12 +\!\frac{c_2}{\sqrt{2\lambda\!-\!c_1^2}}\!
\int^{\sqrt{(2\lambda\!- c_1^2)(T\!-t)}}_0\!\! \frac{\mathrm e^{\frac{z^2}2}}{\sqrt{2\pi}}dz\Big ]\Big \}.
\end{eqnarray}
If instead  condition $\lambda=r$ holds, we arrive in any case to an explicit formula
with $A_1(t,T)$ the same as in (\ref{case1a1}) and $A_2(t,T)$ given by
\begin{equation}
\label{case4a2}
A_2(t,T)=  \lambda\mathrm e^{-\lambda(T-t)}\Big[(T-t)\cN( c_2\sqrt{T-t})+
\sqrt{\frac { T-t}{2\pi c_2^2}}\mathrm e^{-\frac {c_2^2}2(T-t)}- \frac 1{c^2_2}(\cN(c_2\sqrt{T-t})-\frac 12 )\Big].
\end{equation}

\begin{remark}  {\bf (The Merton market model)}

The above results can be extended  with moderate computational effort to the Merton jump-diffusion market model (see e.g. \cite{CC12})
$$
\frac{d S_t}{S_{t-}} = (r-\nu \kappa) dt + \sigma
dW_t + (e^J-1) dN_t,
$$
where $W_t$ is a Brownian motion, $N_t$ a Poisson process with jump-arrival intensity $\nu$, $J \sim \cN(\mu,\delta^2)$ all independent of each other, $\kappa = \E^Q[e^J]-1$ is the jump-size expectation, $\sigma$ is the volatility and $r$ is the risk-free rate.

Under  (IND), for $\alpha=1$  and under the martingale $Q$, if $\tau\sim\exp(\lambda)$,  with $\lambda>0$,
the price of a ({\bf{RTFS}}) option is
given by
\begin{eqnarray}\label{RTFSpriceM}
c(t,T)&=& \mathbf 1_{\{0< \tau\le t\}}call^{M}(t, S_t, S_\tau,T)+
\mathbf 1_{\{\tau>t\}}S_t\sum_{n=0}^{+\infty} \frac {\bar \nu^n}{n!}\left(A_{1,n}(t,T) -
A_{2,n}(t,T)\right),
\end{eqnarray}
where $\bar \nu = \nu (1+\kappa)$, $A_{1,n}(t,T)$ and  $A_{2,n}(t,T)$ are given as in (\ref{A1A2BS}) for  appropriate sets of parameters varying with $n$. The explicit formula is available  from the authors, upon request.
\end{remark}

\bigskip
\noindent\textbf{Pricing with Fourier Transform: the variance Gamma model}

\ms

Here we assume that the price dynamics is of the form $S_T = S_u \mathrm{e}^{X_T-X_u},\, u\le T$, where $X_u$ is a L\`{e}vy process. Fourier transform  is  a well established computational tool for derivative pricing, first introduced in \cite{Hes} and then further developed by several authors (e.g. see \cite{Lee} or   \cite{EGP} and the references therein). Different  ways of applying this technique lead to different representation formulas for the risk neutral price of a plain vanilla call option, we choose to consider
\begin{equation}
call(u,S_u,K,T) = \mathrm{e}^{-r(T-u)} K \frac{1}{2 \pi} \int_{\mathrm{i} \nu - \infty}^{\mathrm{i} \nu + \infty} \mathrm{e}^{- \mathrm{i} z (\log(S_u/K) + r(T-u))} \frac{\phi_{X_T- X_u}(-z)}{\mathrm{i} z - z^2} dz,
\end{equation}
for $z \in \mathbb C$ and $\nu = Im(z) > 1$, with
$\phi_{X_T- X_u}(z)= E[\mathrm{e}^{\mathrm{i} z (X_T- X_u)}]$  the generalized characteristic function of $X_T- X_u$. Because of the independence of the increments,  $\phi_{X_T- X_u}$ does not depend on $S_u$, so the model is scale-invariant and, applying (\ref{scale}),  we may conclude that  the price of the standard Forward Starting option with determination time $u$ is
$$
c(t,u,T) = S_t call(u,1,\alpha,T) = S_t \mathrm{e}^{-r(T-u)} \frac{\alpha }{2 \pi} \int_{\mathrm{i} \nu - \infty}^{\mathrm{i} \nu + \infty} \mathrm{e}^{- \mathrm{i} z (-\log(\alpha) + r(T-u))}  \frac{\phi_{X_T- X_u}(-z)}{\mathrm{i} z - z^2} dz.
$$
By inserting this Fourier representation of the Forward Starting call price into the formulas of Proposition 1.1 or Remark \ref{rem1}, we get a double integral representation of our RTFS call price.

As usual, under hypothesis (IND),  setting $\alpha=1$, we obtain
\begin{eqnarray*}
\!\!\!\!\!\!c(t,T)\!\!\!
 &=&\!\!\! \E^{Q}[B(t,T)(S_T - S_{\tau})^+|\cal G_t]\mathbf 1_{\{0< \tau\le t\}} \\
 &+& \mathbf 1_{\{\tau>t\}}  \frac{S_t}{2 \pi} \int_{\mathrm{i} \nu - \infty}^{\mathrm{i} \nu + \infty} \frac 1{\mathrm{i} z - z^2} \Big [ \int_t^T \mathrm{e}^{- r(T-u)(1+ \mathrm{i} z)} \phi_{X_T- X_u}(-z) \lambda_u\mathrm e^{-\int_t^u\lambda_s ds}du \Big ] dz.
\end{eqnarray*}
In some cases the inner integral can be solved in closed form.
As a matter of fact,  if we choose a Variance Gamma model for the price dynamics
$$
X_t = b Y_t+ cW_{Y_t}, \qquad S_T=S_u \mathrm{e}^{(r+\omega)(T-u) + X_T-X_u},
$$
 with $W_t$ a standard Brownian motion and $Y_t$ a Gamma process independent of $W_t$, with parameters 1 and $\mu$ and $\omega=1/\mu \log(1-b \mu - \mu c^2/2)$ (see \cite{CMC98}),   we have
$$
\phi_{X_T- X_u}(z) = \Big (\frac 1 {1-b\mu z + c^2 \mu \frac{z^2}2}
\Big)^{\frac {T-u}\mu}=
\mathrm e^{-\frac 1\mu(T-u)\ln(1-\mathrm{i} b\mu z + c^2 \mu \frac{z^2}2)},
$$
where $\ln \eta=\ln |\eta|+ i \arg (\eta)$,  with $\eta \in \mathbb C$ and $-\pi \le \arg (\eta)\le \pi$, denotes the principal complex logarithm. Taking also (for ease of computation) a constant hazard rate $\lambda$, we may conclude
\begin{eqnarray} \label{RTFSpriceVG}
 \nonumber
c(t,T) \!\!\!&=& \!\!\!\E^{Q}[B(t,T)(S_T - S_{\tau})^+|\cal G_t]\mathbf 1_{\{0< \tau\le t\}}  \nonumber \\
\!\!\! &+ &\!\!\!\mathbf 1_{\{\tau>t\}}  \frac{S_t}{2 \pi}\lambda \mathrm{e}^{-\lambda(T-t)} \int_{\mathrm{i} \nu - \infty}^{\mathrm{i} \nu + \infty}\frac 1{\mathrm{i} z - z^2}\;
\frac {1 -  \mathrm{e}^{-[r (1+\mathrm{i} z)+\frac 1\mu\ln (1-\mathrm{i}b\mu z + c^2 \mu \frac{z^2}2) - \lambda](T- t)}}{r (1+\mathrm{i} z)+\frac 1\mu\ln (1-\mathrm{i}b\mu z + c^2 \mu \frac{z^2}2) - \lambda}dz,
\end{eqnarray}
leading to a computable formula.
It is reasonable that the same arguments might be extended to tempered stable distributions as well as  to the Carr-Geman-Madan-Yor model (see \cite{CGMY}).

\bigskip
\noindent {\bf Pricing in the Heston stochastic volatility market.}

\ms

Assuming  independence,  we can arrive to an explicit formula  also in the Heston stochastic volatility model, given (under the risk neutral measure) by
\begin{eqnarray*}
dS_t& =&S_t( r + \sqrt{\sigma_t}dW^1_t)\\
d\sigma_t &=& \kappa (\theta - \sigma_t) dt + c\sqrt{\sigma_t}( \rho dW^1_t + \sqrt{1 -\rho^2}dW^2_t),
\end{eqnarray*}
where $r$ is a  constant interest rate.
Here $\cF_t$  has to be the natural filtration generated  by the couple $(S, \sigma)$, which is jointly a bidimensional Markov process,  hence the final pricing formula will depend only upon the initial values of both $S$ and $\sigma$.

As before, we start from the expression
\begin{eqnarray}
\label{heston}
\nonumber
c_H(t,T)&=&
 \E^{Q}[B(t,T)(S_T\!-\alpha S_{\tau})^+|\cal G_t]\mathbf 1_{\{0< \tau\le t\}}\\
&+&
\mathbf 1_{\{\tau>t\}}\Big [\!\int_t^T\!\!c_H(t,u, T, S_t, \sigma_t) \lambda_u\mathrm e^{-\int_t^u\lambda_s ds}du+(1\!-\alpha)S_t\mathrm e^{-\int_t^T\lambda_s ds}
\Big ].
\end{eqnarray}
By employing the results in  \cite{Kruse05} (or \cite{Lucic03}), we can find an explicit formula for the forward start option. Indeed by virtue of the scale invariant property verified also by  this model, we have
\begin{equation}
\label{heston1}
c_H(t,u, T, S_t, \sigma_t)  = \int_0^{+\infty} S_t call_H(u, 1, \alpha, \sigma, T) f_{\sigma_u|\sigma_t}(\sigma) d\sigma,
\end{equation}
where $ call_H(u, 1, \alpha, \sigma, T) $ is the standard price of a call option in the Heston model (for details see for instance \cite{Kruse05} Lemma 2.1) and
\begin{eqnarray}
\label{condden}f_{\sigma_u|\sigma_t}(\sigma)&=&\frac {B(u)} 2 \mathrm e^{-\frac{B(u)\sigma+\Lambda(u)}2}\Big ( \frac {B(u)\sigma}{\Lambda(u)}\Big)^{(R/2-1)/2}I_{R/2-1}(\sqrt{\Lambda (u) B(u)\sigma})\\
\label{coeffB}B(u)&=&\frac{4(\kappa-\rho c)}{c^2}(1-\mathrm e^{-(\kappa-\rho c)(u-t)})^{-1}\\
\label{coefflam}\Lambda(u)&=& B(u)\mathrm e^{-(\kappa-\rho c)(u-t)}\sigma_t, \quad R=\frac {4\kappa \theta}{c^2},
\end{eqnarray}
and $I$ is a modified  Bessel function of the first kind.

{\remark The previous method can be adapted to cover also the case of affine stochastic interest rate market model proposed in  \cite{HP},  where the interest rate follows a Hull and White model and the stochastic volatility is a Ornstein-Uhlenbeck process. }

\section{Numerical results}
\label{sec4a}

In this section we report the results of the numerical implementation of some of the models presented before: the Black \& Scholes, the Variance Gamma and the Heston models, with a random time $\tau \sim \exp(\lambda)$   with parameter $\lambda>0$. In all experiments we set $\alpha=1$, $t=0$ and  we assume the independent hypothesis (IND). Analytical or closed-form prices, eventually requiring numerical quadrature, are compared with standard Monte Carlo approximations of the arbitrage free pricing formula (\ref{random}). Sample paths of the underlying  dynamics are evaluated at the random time $\tau$ and the corresponding payoffs are collected to estimate (\ref{random}). In our experiments we simulated $M=1000000$ samples/paths\footnote{Even though we are not employing them here, we remark that variance reduction techniques (control variates, antithetic variates) or more  efficient simulation algorithms (e.g. the Broadie and Kaya exact simulation, as proposed in \cite{BK}) could improve the MC estimates. Furthermore, alternative simulation schemes exploiting Proposition \ref{gener1} and Remark \ref{rem1} can be easily designed.}.

Adaptive Gauss-Lobatto quadrature was used to approximate  representation integrals (VG and Heston models), which required the evaluation of  extended Fourier transforms involving multivalued functions, such as the complex logarithm. To avoid the well-known numerical instabilities due to a wrong formulation of the characteristic function, we used the so-called \textit{rotation count algorithm} by Kahl and J\"{a}ckel (see e.g. \cite{LK10}). All routines were coded and implemented in MatLab\copyright, version 8.0 (R2012b) on an Intel Core i7 2.40 GHz machine running under Windows 7 with 8 GB physical memory.

For each model we report the closed-form (CF) and the MC prices for different values of $\lambda$, fixing  the other model parameters.  In parenthesis  we report the absolute error w.r.t. the CF price, followed by the corresponding length of the $95 \%$ confidence interval. Without loss of generality we set $S_0=100$, $T=2$ and $r=0$.

In the Black \& Scholes model the RTFS price is given by formulas (\ref{GFSpriceBS})-(\ref{case4a2}). Exact simulations of the geometrical Brownian motion with $\sigma=0.2$ at the random time $\tau$ are used to estimate the pricing formulas.

\begin{table}[h]
  \centering
  \begin{tabular}{c|c|cc}
    \hline
    & CF Price & MC & C.I. Length \\ \hline
   $\lambda=0.25$ & $3.0989$ &  $3.0998$ ($8.4\times 10^{-4}$) & $3.7 \times 10^{-2}$ \\
   $\lambda=0.75$ & $6.6457$ &  $6.6512$ ($5.4\times 10^{-3}$) & $5.5 \times 10^{-2}$ \\
   $\lambda=1.25$ & $8.3710$ &  $8.3699$ ($1.1\times 10^{-3}$) & $6.1 \times 10^{-2}$ \\
   $\lambda=1.75$ & $9.2709$ &  $9.2872$ ($1.6 \times 10^{-2}$) & $6.5 \times 10^{-2}$\\
    \hline
  \end{tabular}
  \caption{\textbf{Black \& Scholes} model for different values of $\lambda$.}\label{TabBS1}
\end{table}

The RTFS prices in the VG model are given by formula (\ref{RTFSpriceVG}). In this example we consider the same parameters as in \cite{CMC98}, $b=-0.1463$, $c=0.1213$ and $\mu=0.1686$. Sample paths are simulated as a Gamma time-changed Brownian motion with a discretization step equal to $1/\sqrt{M}$.

\begin{table}[h]
  \centering
  \begin{tabular}{c|c|cc}
    \hline
    & CF Price & MC & C.I. Length \\ \hline
   $\lambda=0.25$ & $2.0159$ &  $2.0163$ ($4.1\times 10^{-4}$)  & $5.5 \times 10^{-2}$ \\
   $\lambda=0.75$ & $4.3394$ &  $4.3426$ ($3.2\times 10^{-3}$)  & $4.1 \times 10^{-2}$ \\
   $\lambda=1.25$ & $5.4801$ &  $5.4748$ ($5.3\times 10^{-3}$)  & $3.9 \times 10^{-2}$ \\
   $\lambda=1.75$ & $6.0796$ &  $6.0857$ ($6.1 \times 10^{-3}$) & $5.8 \times 10^{-2}$\\
    \hline
  \end{tabular}
  \caption{\textbf{Variance Gamma} model for different values of $\lambda$.}\label{TabVG1}
\end{table}

In Heston's stochastic volatility model parameters  are chosen as in \cite{Kruse05}:  $\sigma_0=0.09$, $\kappa=4$, $\theta=0.06$, $c=0.65$ and $\rho=-0.9$. We need to compute numerically formula (\ref{heston}), which requires evaluating the function given by formula  (\ref{heston1}). This function, though  theoretically well defined for all values of $u$, generates a numerical singularity in the computations, that we treat by approximating
$c_H(0,u,T,S_0,\sigma_0) $ with  $S_0 call_H (u,1,\alpha,\E(\sigma_u),T)$. This gives some advantage since we explicitly have $\E(\sigma_u) = \sigma_0 e^{-\kappa u} + \theta (1-e^{-\kappa u})$. Comparison with Monte Carlo approximations is run with  sample paths  simulated by the basic Euler scheme with a step length equal to $\sqrt{M}$.

\begin{table}[h]
  \centering
  \begin{tabular}{c|c|cc}
    \hline
    & CF price & MC & C.I. Length   \\ \hline
   $\lambda=0.25$ & $3.4907$  & $3.4903$ ($3.7 \times 10^{-4}$) & $3.6 \times 10^{-2}$   \\
   $\lambda=0.75$ & $7.5290$  & $7.5227$ ($6.2 \times 10^{-3}$) & $5.2 \times 10^{-2}$   \\
   $\lambda=1.25$ & $9.5307$  & $9.5157$ ($1.4 \times 10^{-2}$) & $5.7 \times 10^{-2}$   \\
   $\lambda=1.75$ & $10.5988$ & $10.5920$ ($6.8 \times 10^{-3}$)& $6.0 \times 10^{-2}$   \\
    \hline
  \end{tabular}
  \caption{\textbf{Heston} model for different values of  $\lambda$.}\label{TabHes1}
\end{table}

\medskip

\begin{figure}[t]
\begin{center}
\hspace{-2.1cm}
\includegraphics[width=18cm,height=6cm]{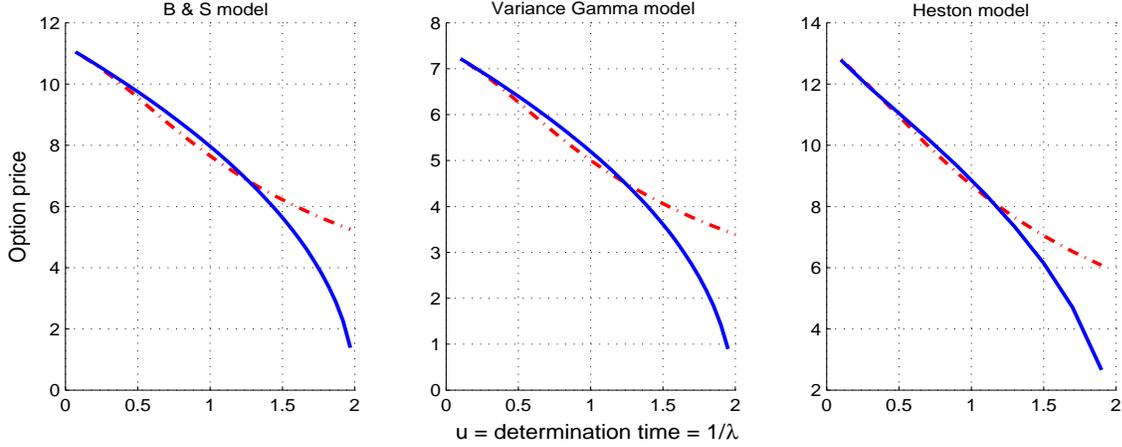}
\caption{\small Prices of RTFS (red dash-dotted lines) and FS (blue solid lines) options in the Black \& Scholes,  Variance Gamma and  Heston model as a function of the expectation of the random time $\tau$. \label{FigAll1}}
\end{center}
\end{figure}

Lastly, we choose  the random time $\tau \sim \exp(\lambda)$ with varying  $\ds\lambda \ge \frac 12$ in order to  compare  at time $t=0$,   the RTFS price and the  price of a FS option with determination time $u=\E(\tau)=\frac 1 \lambda \leq T$. In other words we compare $\E((S_T- S_{\tau\wedge T})^+)$ with $ \E((S_T- S_{\E(\tau)})^+)$, see Figure (\ref{FigAll1}). In all the instances prices decrease as $u$ increases, moreover we notice that: i) for $u \rightarrow 0$ (i.e. $\lambda \rightarrow +\infty$) FS and RTFS prices both tend to the price of an ATM plain vanilla call; ii) for $u \rightarrow T$ (i.e. $\lambda \rightarrow 1/2$) the FS value trivially goes to $0$, while the value of a RTFS option stays positive since $Q(\tau \leq T) \rightarrow 1-1/e \approx 63 \%$.

\section{Counterparty credit risk for RTFS options}
\label{sec5a}

In this section we are interested in computing the credit value adjustment  (CVA) for a RTFS option, when one of the parties is subject to a possibly additional default time.

The interest lies in the fact that those products would be  OTC and knowing that the CVA is computable might be convenient.

We first recall what the CVA is for an additive cashflow.  Let us denote by $\Pi(t, T)$ the discounted value of the cashflow between   time  $t$  and $T$; this value has to be additive, which means that for   times $t\le s\le u$ it holds
\begin{equation}
\label{additive}
\Pi(t,s ) + B(t,s)\Pi(s, u)= \Pi (t,u).
\end{equation}
All European contingent claims with integrable, $\cG_T-$measurable payoff $h$,    trivially verify  (\ref{additive}) since
$
\Pi(t, s)= h \mathbf 1_{\{s=T\}}$.
We call Net Present Value  of $\Pi$
\begin{equation}
\label{NPV}
\textrm{NPV}(t)= \E( \Pi(t,T)|\cG_t).
\end{equation}
We now look into unilateral counterparty credit risk.
Let us suppose that the cashflow is between two parties  $B$ (buyer) and  the counterparty $C$, that is subject to default, with default time $\tau_C$. Then, the discounted value of the cashflow  has to be adjusted by subtracting from the defaultless case the quantity
\begin{equation}
\label{CVA}
{\textrm {CVA}}(t)=(1-R)\E(\mathbf 1_{\{t<\tau_C\le T\}}B(t, \tau_C)\textrm{ NPV}(\tau_C)^+|\cG_t)
\end{equation}
where $0 \le R\le 1 $ represents a deterministic recovery rate (see \cite{BC}).

In the case of classical Forward Start options with fixed strike-determination time $u$, it is possible to compute directly the CVA. We denote by  $\{\cH^C_t\}$ the filtration generated by the process $H^C_t:= \mathbf 1_{\{\tau_C\le t\}}$ and by $\cG^C_t = \cF_t\lor \cH^C_t$.  Under  independence between $\{\cF_t\}$ and $\{\cH^C_t\}$, taking for granted that $B(t,u)S_u$ remains a  martingale in the enlarged filtration $\{\cG_t^C\}$ w.r.t.  a risk neutral probability measure  $Q^C$ which extends the risk neutral probability $P$, we have
\begin{equation}
\label{CVAFS}
\textrm{CVA}^{FS}(t)= (1-R)Q^C(t<\tau_C<T) \E^{Q^C}(B(t,T)(S_T- \alpha S_u)^+|\cF_t),
\end{equation}
where the key Lemma and  the same notation as in section 1 were used, under the assumption that $Q^C(\tau_C>t)=1$.

To extend the previous formula to the case of  the RTFS options we need to consider two random times: the strike-determination time $\tau $ and the counterparty default time $\tau_C$.

Hence  we need to extend the original filtration $\{\cF_t\}$ with both $\{\cH_t\}$ and $\{ \cH^C_t\}$, obtaining
$$
\cG_t= \cF_t\lor \cH_t\lor  \cH^C_t.
$$
Accordingly, we  assume there exists an extension  of the risk neutral probability $P$ to a common probability  space $(\Omega, \cG_T)$ that we still denote by $Q$ and we set the corresponding key assumption

\medskip

\noindent
{\bf (HC)}\qquad \qquad\qquad \qquad  Every $\cF_t$ martingale remains a $\cG_t$ martingale.\hfill

\medskip

Taking  $\alpha=1$ for simplicity, we have that the price of a defaultable RTFS  call option is
$$
\bar c(t,T):=\E^{Q}[\mathbf 1_{\{\tau_C>T\}}B(t,T)(S_T- S_{\tau\wedge T})^+|\cal G_t].
$$
So, in presence of a recovery rate $R$, we obtain that the CVA for this product has to be
\begin{eqnarray}\label{randomdef}
\!\!\!\!\!\textrm{CVA}(t,T)\!\!\!&:=&\!\!\!
(1-R)[
c(t,T)- \bar c(t,T)]= (1\!-\!R) \E^{Q}[\mathbf 1_{\{t<\tau_C\le T\}}B(t,T)(S_T-S_{\tau\wedge T})^+|\cal G_t].
\end{eqnarray}
We may decompose the expectation as
\begin{eqnarray*}
\!\!\!&&\!\!\!\E^{Q}[\mathbf 1_{\{t<\tau_C\le T\}}B(t,T)(S_T-S_{\tau\wedge T})^+|\cal G_t]\\
\!\!\!&=&\!\!\!\E^{Q}[\mathbf 1_{\{t<\tau_C\le T\}}\mathbf 1_{\{\tau> T\}}B(t,T)(S_T\!-\!S_{\tau\wedge T})^+|\cal G_t]\!+\E^{Q}[\mathbf 1_{\{t<\tau_C\le T\}}\mathbf 1_{\{\tau\le T\}}B(t,T)(S_T\!-\!S_{\tau\wedge T})^+|\cal G_t]\\
\!\!\!&=&\!\!\!\E^{Q}[\mathbf 1_{\{t<\tau_C\le T\}}\mathbf 1_{\{t<\tau\le T\}}B(t,T)(S_T-S_{\tau})^+|\cal G_t]
+\mathbf 1_{\{\tau\le t\}}\E^{Q}[\mathbf 1_{\{t<\tau_C\le T\}}B(t,T)(S_T-S_{\tau})^+|\cal G_t],
\end{eqnarray*}
where the last equality is justified by the fact that the first term in the first passage is equal to 0. In the second equality the crucial term is the first, as the second reduces to the CVA of  a standard forward start option, for which formula (\ref{CVAFS}) applies.
We may handle the first term under condition of independence between $\tau_C$ and $(\tau,   \{S_t\})$.
Denoting  by $\cF^H_t=\cF_t \lor \cH_t$, we have $\cG_t = \cF^H_t\lor \cH^C_t$ and applying the key lemma to $\cG_t$ and $\cF^H_t$ we obtain
\begin{eqnarray*}
&&
\E^{Q}[\mathbf 1_{\{t<\tau_C\le T\}}\mathbf 1_{\{t<\tau\le T\}}B(t,T)(S_T-S_{\tau})^+|\cal G_t]\\
&=&Q(\tau_C>t|\cG_t)\frac{\E^{Q}[\mathbf 1_{\{t<\tau_C\le T\}}\mathbf 1_{\{t< \tau\le T\}}B(t,T)(S_T-S_{\tau})^+|\cF^H_t]}{Q(\tau_C>t|\cF^H_t)}.
\end{eqnarray*}
The independence of $ \mathbf 1_{\{t<\tau\le T\}}$ from the remaining factors gives
\begin{eqnarray*}
&&
\E^{Q}[\mathbf 1_{\{t<\tau_C\le T\}}\mathbf 1_{\{t<\tau\le T\}}B(t,T)(S_T-S_{\tau})^+|\cal G_t]\\
&=&Q(\tau_C>t|\cG_t)\frac{ Q(t< \tau_C\le T)}{Q(\tau_C>t)}\E^{Q}[\mathbf 1_{\{t< \tau\le T\}}B(t,T)(S_T-S_{\tau})^+|\cF^H_t]
\end{eqnarray*}
and if   $Q( \tau_C>t)=1$, the above reduces to
$$
\E^{Q}[\mathbf 1_{\{t<\tau_C\le T\}}\mathbf 1_{\{t<\tau\le T\}}B(t,T)(S_T-S_{\tau})^+|\cal G_t]=Q(t< \tau_C\le T) \E^{Q}[\mathbf 1_{\{t< \tau\le T\}}B(t,T)(S_T-S_{\tau})^+|\cF^H_t]
$$
which is the weighted value of a RTFS option, exactly as it happened with the standard FS option.
Summarizing we have

\begin{proposition} Under hypothesis (HC)  and the condition that $Q(\tau_C >t)=1$, if  $\tau_C$ is independent of
$(\tau,   \{S_t\})$, then the CVA of a defaultable RTFS call option of price $c(t,T)$, with recovery rate $R\in [0,1]$,  is given by
$
\textrm{CVA}(t,T)=(1- R)Q(t< \tau_C\le T) c(t,T)$.

\end{proposition}

Lastly we remark that if the two random times coincide, $Q(\tau=\tau_C)=1$, then we may conclude that
$$
CVA(t,T)=(1- R) c(t,T),
$$
which implies that  the defaultable price is  $\bar c(t,T)=R c(t,T)$.

\section*{Acknowledgements}
The authors  thank Prof. C. Chiarella for useful suggestions at  an early stage of this work.


\begin{thebibliography}{999999}


\bibitem{AR} R. Ahlip, M. Rutkowski, \emph{Forward start options under stochastic volatility and stochastic interest rates}, International Journal of Theoretical and Applied Finance, Vol. 12, No. 2, 209-225, (2009).

\bibitem{AM09} C. Albanese, H. Lo, A. Mijatovic, \emph{Spectral methods for volatility derivatives}, Quantitative Finance, Vol. 9, No. 6,  663-692, (2009).

\bibitem{AN07} C. Alexander, L.M. Nogueira \emph{Model-free hedge ratios and scale invariant models}, Journal of Banking and Finance, Vol. 31, No. 6, 1839-1861, (2007).

\bibitem{BCB} T. R. Bielecki, S. Crepey, D. Brigo, \emph{Counterparty Risk and Funding: A Tale of Two Puzzles}, Chapman and Hall/CRC , (2014).

\bibitem{BJR} T. R. Bielecki, M. Jeanlanc, M. Rutkowski, \emph{Valuation and Hedging of Credit Derivatives} Lecture notes CIMPA- UNESCO Morocco School (2009).

\bibitem{BR} T. R. Bielecki, M. Rutkowski, \emph{Credit Risk: Modeling, Valuation and Hedging}, Series: Springer Finance (2002).

\bibitem{BC} D. Brigo, K .Chourdakis, \emph{Counterparty risk for credit default swaps: impact of default volatility and spread correlation}, International Journal of Theoretical and Applied Finance, Vol. 12, No. 7, 1007-1029, (2009).

\bibitem{BK} M. Broadie,O. Kaya, \emph{Exact Simulation of Stochastic Volatility and other Affine Jump Diffusion Processes}, Operations Research, Vol. 54, No. 2, 217-231, (2006).

\bibitem{CGMY} P. Carr, H. Geman, D. Madan, M. Yor, \emph{The fine structure of asset returns: an empirical Investigation}, Journal of Business, Vol. 75, No. 2, 305-332, (2002).

\bibitem{CM} P. Carr, D. Madan, \emph{Option Valuation Using the Fast Fourier Transform}, Journal of Computational Finance, Vol.  2, 61-73, (1999).

\bibitem{CMC98} P. Carr, D. Madan, E. Chang, \emph{The Variance Gamma Process and Option Pricing}, European Finance Review 2, 79�105, (1998).

\bibitem{CC12} G. H. Cheang, C. Chiarella, \emph{A Modern View on Merton's Jump-Diffusion Model}, in "Stochastic Processes, Finance and Control: A Festschrift in Honor of Robert J. Elliott", Samuel N. Cohen, Dilip Madan, Tak Kuen Siu and Hailiang Yang (eds.), pg 217-234, World Scientific, (2012).

\bibitem{CCZ} G.H. Cheang, C. Chiarella, A. Ziogas, \emph{ The representation of American options prices under stochastic volatility and jump-diffusion dynamics},  Quantitative Finance, Vol. 13,  No. 2, 241-253, (2013).

\bibitem{CWW}  C-Y. Chen, H-C. Wang, J-Y. Wang, \emph{ The valuation of forward-start rainbow options}, Review Derivative Research, DOI 10.1007/s11147-014-9105-0, (2014).

\bibitem{DM}  C. Dellacherie - P. A. Meyer \emph{ Probabilities and Potentials B}, North Holland (1982).

\bibitem{EGP} E. Eberlein, K. Glau, A. Papapantoleon, \emph{Analysis of Fourier transform valuation formulas and applications}, Applied Mathematical Finance, Vol. 17, n.3, 211-240, (2010).

\bibitem{GJ08} P. V. Gapeevy, M. Jeanblanc, {\emph On filtration immersions and credit events}, CDAM Research Report LSE-CDAM-2008-24, (2008).

\bibitem{GW99}, S. F. Gray, R. E. Whaley, \emph{Reset put options: valuation, risk characteristics, and an application}, Australian Journal of Management, Vol. 24, 1,1-20, (1999).

\bibitem{GH08} J. H. Guo, M. W. Hung, \emph{A generalization of Rubinstein's "Pay Now, Choose Later"}, The Journal of Futures Markets, Vol. 28, No. 5, 488-515 (2008).

\bibitem{HP} A. van Haastrecht, A. Pelsser \emph{Accounting for Stochastic Interest Rates, Stochastic Volatility and a General Correlation Structure in the Valuation of Forward Starting Options}, Journal of Futures Markets, Vol. 31, No. 2, 103-125 (2011).

\bibitem{Hes} S. L. Heston, \emph{Closed-Form Solution for Options with Stochastic Volatility with Applications to Bond and Currency Options}, Rev. Fin. Studies,  vol. 6, 327-343, (1993).

\bibitem{Lee} R. W. Lee,  \emph{Option pricing by transform methods: extensions, unification and error control}, Journal of Computational Finance, Vol. 7, n. 3, 51-86, (2004).

\bibitem{LK10} R. Lord and C. Kahl, \emph{Complex logarithm in Heston-like models},  Mathematical Finance, Vol. 20, No. 4, 671-694, (2010).

\bibitem{Lucic03} V. Lucic, \emph{Forward-start Options in Stochastic Volatility Models}, Wilmott magazine, (2003).

\bibitem{Kruse05} S. Kruse and U. N\"{o}gel, \emph{On the pricing of forward starting options in Heston's model on stochastic volatility}, Finance and Stochastics, Vol. 9, 233-250,  (2005).

\bibitem{P} P. Protter  \emph{Stochastic Integration and Differential Equations}, Springer-Verlag Berlin, (2004).

\bibitem{Ram12} A. Ramponi, \emph{On Fourier Transform Methods for Regime-Switching Jump-Diffusions and the pricing of Forward Starting Options}, International Journal of Theoretical and Applied Finance, Vol. 15, No. 5,  (2012).

\bibitem{Rubin} M. Rubinstein, \emph{Pay now choose later}, Risk, Vol. 44, 44-47, (1991).

\bibitem{Wilmott} P. Wilmott, \emph{Cliquet options and volatility models},  Wilmott Magazine, 78-83, (2002).

\end{thebibliography}
\end{document}